\documentclass[12pt,preprint]{aastex}
\newcommand{\mjyb}{\,mJy~beam$^{-1}$}
\newcommand{\etal}{~et al.}

\shorttitle{The HD\,107146 Debris Disk}
\shortauthors{Corder et al.} 

\begin{document}

\title{A Resolved Ring of Debris Dust around the Solar Analog HD\,107146}

\author{Stuartt Corder\altaffilmark{1,2}, 
        John M. Carpenter\altaffilmark{1}, 
        Anneila I. Sargent\altaffilmark{1}, 
        B. Ashley Zauderer\altaffilmark{3},
        Melvyn C.~H. Wright\altaffilmark{4}, 
        Stephen M. White\altaffilmark{3},
        David P. Woody\altaffilmark{5},
        Peter Teuben\altaffilmark{3}, 
        Stephen L. Scott\altaffilmark{5}, 
        Marc W. Pound\altaffilmark{3}, 
        Richard L. Plambeck\altaffilmark{4}, 
        James W. Lamb\altaffilmark{5}, 
        Jin Koda\altaffilmark{1},
        Mark Hodges\altaffilmark{5}, 
        David Hawkins\altaffilmark{5}, 
        and
        Douglas C.-J. Bock\altaffilmark{6}
       }

\altaffiltext{1}{Division of Physics, Mathematics and Astronomy,
California Institute of Technology, MS 105-24 Caltech, Pasadena, CA
91125}

\altaffiltext{2}{Jansky Fellow, Current Address Av. Apoquindo 3650, Piso
18, Las Condes, Santiago, Chile}

\altaffiltext{3}{Department of Astronomy, University of Maryland,
College Park, MD 20742, USA}

\altaffiltext{4}{Department of Astronomy and Radio Astronomy
Laboratory, University of California, Berkeley, CA 94720, USA}

\altaffiltext{5}{Owens Valley Radio Observatory, California Institute
of Technology, P. O. Box 968, Big Pine, CA 93513, USA}

\altaffiltext{6}{Combined Array for Research in Millimeter-wave
Astronomy, P.O. Box 968, Big Pine, CA 93513, USA}

\begin{abstract}
We present resolved images of the dust continuum emission from the
debris disk around the young (80-200\,Myr) solar-type star HD\,107146
with CARMA at $\lambda$1.3\,mm and the CSO at $\lambda$350\,\micron.
Both images show that the dust emission extends over an $\sim$10$''$
diameter region. The high resolution (3$''$) CARMA image further
reveals that the dust is distributed in a partial ring with
significant decrease in flux inward of 97\,AU. Two prominent emission
peaks appear within the ring separated by $\sim$140\arcdeg\ in
position angle.  The morphology of the dust emission is suggestive of
dust captured into a mean motion resonance, which would imply the
presence of a planet at an orbital radius of $\sim$45-75\,AU.
\end{abstract}


\section{Introduction}
\label{intro}

Photometric surveys with {\it IRAS}, {\it ISO}, and the {\it Spitzer} 
Space Telescope have identified hundreds of main-sequence 
stars where the observed flux densities at mid-infrared wavelengths are 
brighter than the stellar photosphere \citep[for a review, see][]{Meyer07}.
This ``excess'' infrared emission is generally attributed to small dust
grains that are produced when planetesimals collide and fragment
into smaller debris. 

The gravitational interaction of any planets with the debris dust produces gaps and
peaks in the dust spatial distribution that provides clues to the underlying
planetary system \citep{Liou99}. However, the dust emission in most debris
systems remains unresolved, and the dust spatial distribution is often inferred
by modeling the composite spectral energy distribution. In practice
such analyses yield ambiguous results, since simplified disk geometries are
assumed, and the inferred dust spatial distribution depends sensitively on the
assumed grain composition, porosity, and size.

A few nearby debris disks have been resolved, providing direct information
on the dust spatial distribution. Scattered light images at visible and 
near-infrared images often show smooth azimuthal distributions, frequently
distributed in ring-like geometries \citep{Kalas04,Kalas05,Schneider06}.
Resolved millimeter wavelength images also show rings of
material, but with more clumpy distributions of debris emission
\citep{Holland98,Greaves98,Greaves05,Koerner01,Wilner02,Maness08}. 
These clumps have been interpreted as signposts of planetary systems where the
gravitational interactions trap the dust in mean motion resonances
\citep{Wilner02,Wyatt03}.

The debris disk around the G2V star HD\,107146 \citep{Jaschek78} has
received special interest since it is a rare example of a resolved debris
disk around a young \citep[80--200\,Myr;][]{Williams04} solar analog.
At a distance of 28.5\,pc \citep{Perryman97}, the disk has been 
resolved in near-infrared scattered light \citep{Ardila04} and in the
millimeter thermal continuum \citep{Williams04,Carpenter05}. We present new
(sub-)millimeter-wavelength images of the HD\,107146 debris disk obtained with
CARMA (Combined Array for Research in Millimeter-wave Astronomy) and the CSO
(Caltech Submillimeter Observatory). The debris disk is clearly resolved in 
both datasets, with the CARMA image revealing a complex morphology not seen 
previously. We describe the observations in \S\ref{obs}, present the
resolved images in \S\ref{images}, and discuss the implications of these
results in \S\ref{implications}. 

\section{Observations}\label{obs}

Continuum images of HD\,107146 were obtained at wavelengths of 1.3\,mm
and 350\,\micron\ using CARMA and the CSO, respectively. We adopted
equatorial coordinates of ($\alpha$,$\delta$) =
(12:19:06.50,+16:32:53.87) for equinox and epoch J2000, with a proper
motion of ($\mu_\alpha$,$\mu_\delta$) =
($-$175.65,$-$148.28)\,mas~yr$^{-1}$ \citep{Perryman97}.

CSO observations were carried out between UT April 17 and 21, 2005,
using the Submillimeter High Angular Resolution Camera II
\citep[SHARC~II;][]{Dowell03} with the 350\,\micron\ filter. The
median zenith opacity at 225\,GHz was 0.04, which equates to an
opacity of $\sim$1 at 350\,\micron. Telescope pointing was checked on
Callisto every 30~minutes, which was 20\arcdeg\ away from HD\,107146
at the time of the observations. Pointing was also checked on the
fainter source 3C273, which is 15\arcdeg\ from HD\,107146, every
hour. The pointing offsets as a function of time and elevation were
interpolated in offline data processing and applied before co-adding
the HD\,107146 data. Absolute flux calibration was set from
observations of Neptune assuming a total flux density of 92.6\,Jy. We
estimate a 1$\sigma$ calibration uncertainty of 15\% from repeated
observations of calibrators over the entire observing run.  Images
were produced using the Comprehensive Reduction Utility for SHARC~II
\citep[CRUSH;][]{Kovacs08}. The ``deep'' data reduction mode in CRUSH
was applied to both the calibrators and HD\,107146.

CARMA observations were conducted on UT 2007 August 21, 2008 September
12, and 2008 September 14 in E-configuration, and on four days between
UT March 3 and March 7, 2008 in D-configuration.  The E- and D-
configurations span baselines of 6--66\,m and 11--148\,m,
respectively.  Observations were conducted with the local oscillator
(LO) set to a frequency of 227.25\,GHz. Three correlator bands of
468\,MHz bandwidth each were placed at intermediate frequencies (IF)
of 2.25, 2.75, and 3.25\,GHz to provide a total continuum bandwidth of
2.8\,GHz after combining the upper and lower sidebands.  The quasar
3C273 was observed approximately every 20 minutes for amplitude,
phase, and passband calibration. For the March and September 2008
observations, pointing was updated every $\sim$30 minutes using
optical reference pointing (Corder, Carpenter, \& Wright~2008, in
preparation).

All reduction and imaging were performed using MIRIAD \citep{Sault95}.
Data were flux calibrated by observing Mars or MWC\,349. For MWC\,349, we
assumed a flux density of 1.69\,Jy following the calibration adopted at the
PdBI\footnote{http://www.iram.fr/IRAMFR/IS} with an assumed uncertainty of 10\%
\citep{Altenhoff94}. The night-to-night RMS repeatability of the measure flux 
density of 3C273 is 2.4\%. We adopt a net calibration uncertainty of 15\%.
The image was formed with natural weighting of the $uv$ visibilities and a 
Gaussian taper of FWHM=1.6\arcsec, and then deconvolved with the point spread 
function (i.e. the ``dirty'' beam) using a hybrid H\"ogbom/Clark/Steer 
algorithm \citep{Hogbom74,Clark80,Steer84}. Since the emission is contained
to within a radius of $<$5$''$ from the phase center, no primary beam corrections
were applied. The final image has a resolution of 3.2$''\times$2.7$''$ and a 
RMS noise of 0.35\mjyb.

\section{Resolved Images}
\label{images}

\begin{figure}
\hspace{-0.5cm}
\includegraphics[angle=-90,scale=0.7]{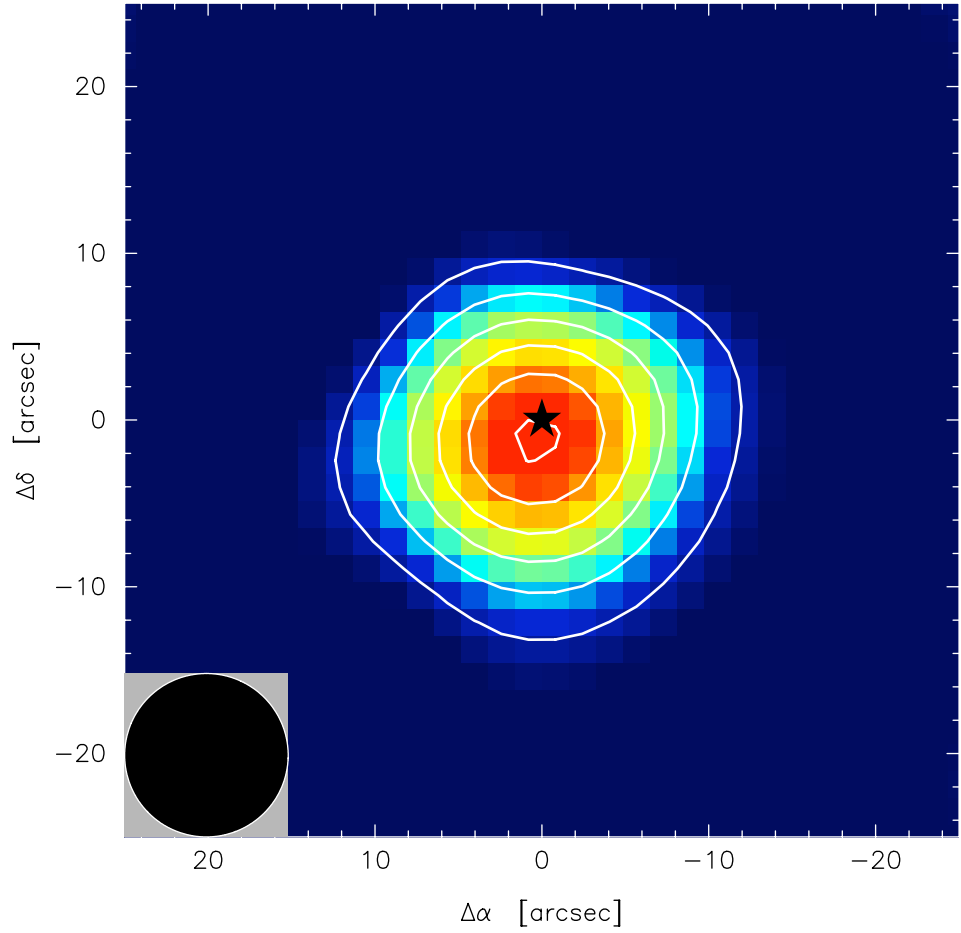}
\caption{
  \label{fig:cso} 
  CSO 350\,\micron\ continuum image of the HD\,107146 debris disk.
  Contours begin at 3$\sigma$ with increments of 5$\sigma$, where
  $\sigma=6$\mjyb. The filled star at (0,0) indicates the stellar position. 
  The filled circle in the lower left corner represents the 9.8\arcsec\ 
  resolution of the observations measured on the Callisto image.
}
\vspace{0.40cm}
\end{figure}

Figure~\ref{fig:cso} presents the CSO 350\,\micron\ continuum image of HD\,107146.
The RMS noise in the image is 6\mjyb, and the peak flux density (181\mjyb) is
detected with a signal to noise ratio of 30. The integrated 350\,\micron\ flux
density obtained by fitting an elliptical Gaussian to the image is
319$\pm$45\,mJy, where the $1\sigma$ uncertainty includes the statistical
(6\,mJy) and calibration (45\,mJy) uncertainties added in
quadrature. The centroid of the 350\micron\ continuum emission is offset from
the stellar position by ($\Delta\alpha,\Delta\delta$) = (0.5$''$,-1.0$''$); 
this offset is consistent within the pointing accuracy of the CSO telescope. No
significant azimuthal structure in the 350\,\micron\ emission is detected.

\begin{figure}
\hspace{2cm}
\includegraphics[angle=0,scale=0.75]{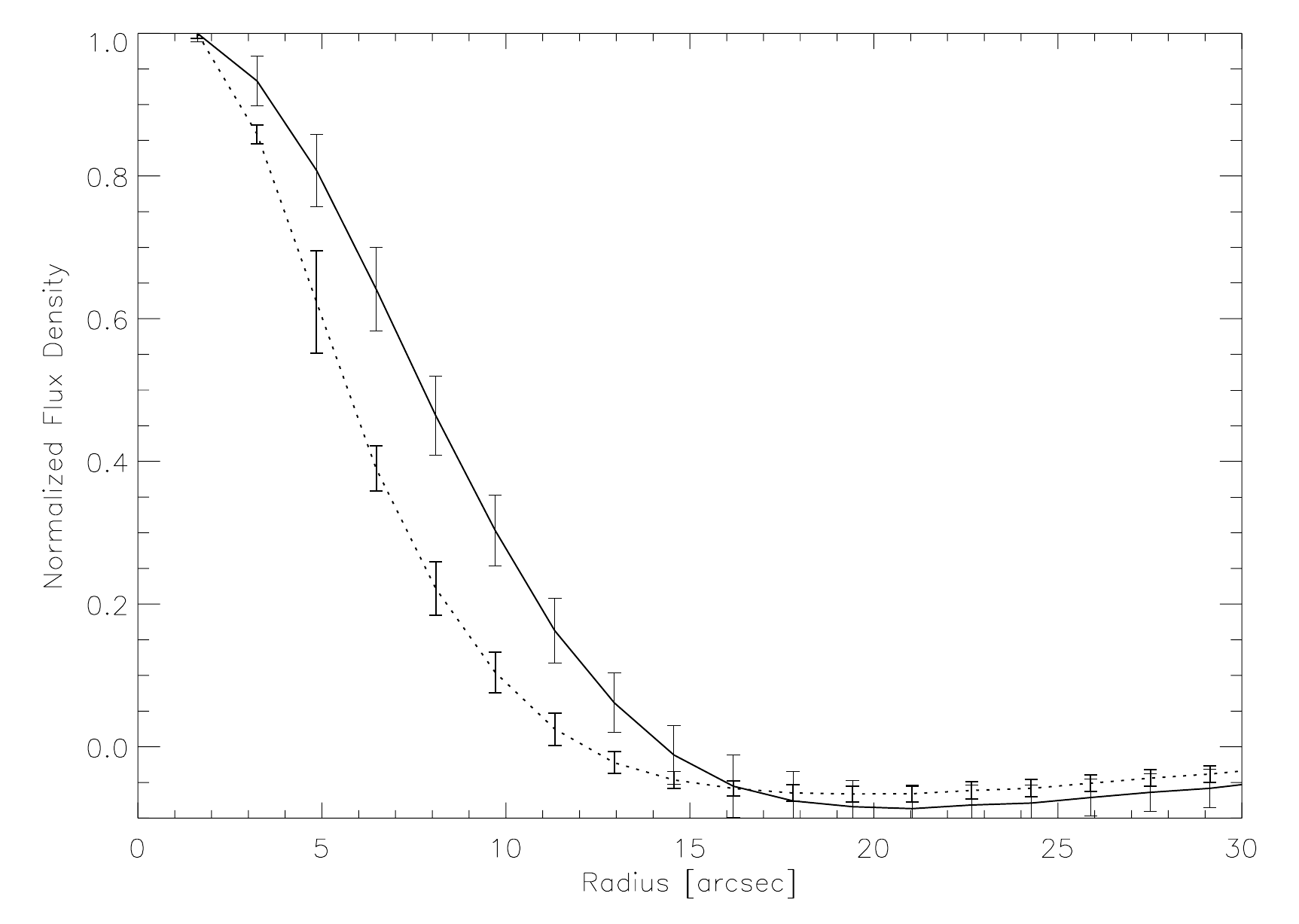}
\caption{
  \label{fig:cso_profile} 
  Radial profile of the CSO 350\,\micron\ continuum emission for the HD\,107146 
  debris disk (solid curve) and Callisto (dashed curve). Both radial profiles 
  were measured on coadded images of the sources.
}
\vspace{0.40cm}
\end{figure}

Figure~\ref{fig:cso_profile} shows the azimuthally-averaged radial
profile of the 350\,\micron\ emission for HD\,107146 (solid curve) and
Callisto (dashed curve). Callisto had an angular diameter of
1.49\arcsec\ at the time of the observations and is considered a point
source for this analysis. The measured FWHM of the emission
profile on Callisto is 9.8\arcsec$\pm$0.03\arcsec.  By contrast, the FWHM
size of the HD\,107146 debris disk measured from an elliptical Gaussian fit 
to the 350\,\micron\ image is $13.5''\times 12.6''$. The deconvolved image
size of HD\,107146 using the observed Callisto image to represent the 
beam profile
is $(8.9''\pm 0.6'') \times (8.2''\pm 0.5'')$. The 350\,\micron\
continuum emission toward HD\,107146 is clearly resolved. 

\begin{figure}
\hspace{-0.5cm}
\includegraphics[angle=-90,scale=0.7]{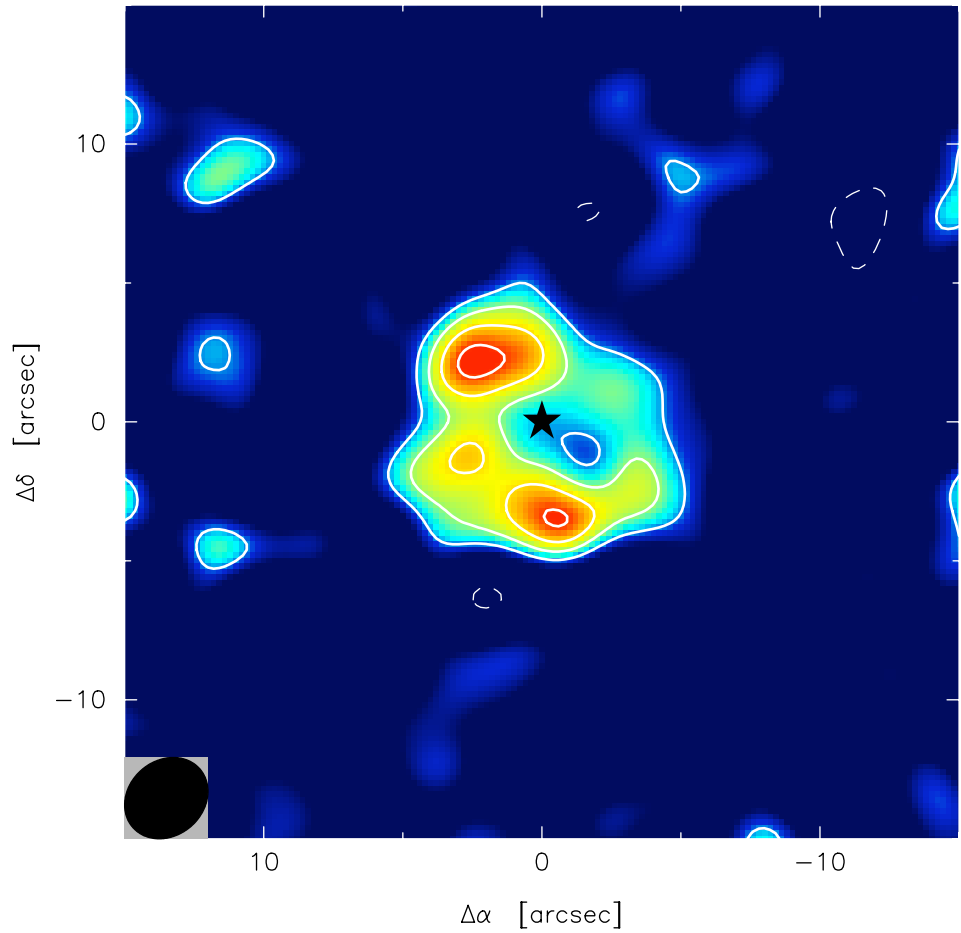}
\caption{
  \label{fig:carma} 
  CARMA 1.3\,mm continuum image of the HD\,107146 debris disk.
  The FWHM of the synthesized beam ($3.2''\times2.7''$) is shown in the
  lower left corner. Solid contours begin at 2$\sigma$ with increments of
  1$\sigma$, where $\sigma = 0.35$\mjyb. Dashed contours begin at $-2\sigma$;
  no $< -3\sigma$ pixel values are present within the region shown. The filled
  star at (0,0) indicates the stellar position. The ``dirty'' beam has been
  deconvolved, but no primary beam corrections have been applied. 
} 
\vspace{0.40cm} \end{figure}

Figure~\ref{fig:carma} presents the CARMA 1.3\,mm continuum image of
HD\,107146 at 3\arcsec\ resolution. The 1.3\,mm continuum emission
extends over a $\sim$10\arcsec\ region.  The integrated flux density
obtained by summing the emission within a $12''\times12''$ box
centered on the stellar position is $10.4\pm1.4$\,mJy, with an
additional 15\% calibration uncertainty (\S\ref{obs}).

The 1.3\,mm emission is resolved into a partial ring that encircles the
stellar position. The locations of the emission peaks are robust to various 
calibration schemes and data reduction approaches, but the magnitude of the 
peaks and the gap in the northwestern part of the ring are somewhat variable
with such changes. The two brightest positions in the ring are located
nearly equidistant from the stellar position on opposite sides of the
star. The northeast peak is located 3.3\arcsec\ from the star at a
position angle (east of north) of PA=48\arcdeg, and the southwest peak
is offset by 3.5\arcsec\ from the stellar position at PA=190\arcdeg. The 
projected distance of the dust peaks from the star is $\sim$97\,AU.

The clumpy structure observed in the CARMA 1.3\,mm continuum image is
in contrast to the smooth, continuous ring observed in scattered light
at optical wavelengths \citep{Ardila04}. Further, the scattered-light image 
peaks at a larger orbital radius (130\,AU) than the dust continuum image. 
The differences in the radial distribution is qualitatively consistent with the
notion that small grains, which are efficient at scattering, are ejected by 
radiation pressure from the system, while the larger grains are
found at smaller radii \citep{Takeuchi01,Wyatt06}.

The morphological differences of the 350\,\micron\ and 1.3\,mm images can be
attributed to resolution in that if we produce the 1.3\,mm CARMA image with a
8\arcsec\ FWHM taper to yield a 9\arcsec\ synthesized beam, the resulting image
is centrally peaked on the stellar position and is similar in size to the
350\,\micron\ image. In detail, other published, resolved, (sub-)millimeter
images are inconsistent with the CSO and CARMA images. \citet{Williams04}
report that the 450\,\micron\ emission at 8\arcsec\ resolution is offset from 
the
stellar position by 4.4\arcsec. Since the 350\,\micron\ (see Fig.~\ref{fig:cso})
and the 850\,\micron\ image \citep{Williams04} are centered on the stellar
position to within 2\arcsec, we assume the offset in the 450\,\micron\ image is a
pointing error. A more significant discrepancy is seen between the CARMA
1.3\,mm image and the OVRO 3\,mm image \citep{Carpenter05}. The 3\,mm emission
was resolved at $4.5'''\times4.0''$ resolution with a FWHM size of
$(6.5''\pm1.4) \times (4.2''\pm1.3'')$. We have smoothed the 1.3\,mm image to
the resolution of the 3\,mm image, and the ring structure is not apparent at
the coarser resolution. However, the apparent ellipticity in the 3\,mm image is
nearly orthogonal to the clumps in the 1.3\,mm image. Since the 1.3\,mm 
image has higher signal to noise, finer resolution, and more baselines for
better imaging, we view that image as a more reliable tracer of the debris
structure.

\section{Implications}
\label{implications}

Assuming gas drag is negligible, the orbital lifetime of dust grains in an
optically thin disk is limited by Poynting-Robertson drag, collision grinding
of particles combined with radiation blowout of small grains, and stellar-wind
drag. \citet{Dominik03} have shown that for debris disks detectable with
current instrumentation, collisional grinding of particles to the radiation
blowout size dominates over Poynting-Robertson drag in removing mass from the
debris system. Stellar-wind drag may also be important \citep{Gustafson94}, 
but the magnitude of this effect is poorly constrained at young ages
\citep{Jura04,Wood05}.

If a collisional cascade has produced the dust particles in the HD\,107146 
debris disk, the expected particle size distribution is $n(a)
\propto a^{-3.5}$ \citep{Dohnanyi69} for grain sizes larger than the radiation
blowout size (radius $\sim$0.5\,\micron\ for a solar-type star). In practice, 
the truncated distribution will produce a ``wavy'' pattern of particle sizes
superimposed on the power-law \citep[e.g.][]{Thebault07}, but we neglect that
complication. For silicate grains \citep{Weingartner01} larger than the
blowout size, half of the flux density at a wavelength of 1.3\,mm will be
emitted by grains larger than $\sim$0.5\,mm in radius for a $a^{-3.5}$ size 
distribution. The ratio of radiation to gravitational forces ($\beta$) for 
these particles is $\beta$$\sim$0.001, indicating radiation pressure has 
negligible effect on their dynamics. 

By the above arguments, the millimeter-wave emission traces predominantly
millimeter-sized particles where the dynamics are controlled by particle
collisions and gravity. This is consistent with clumpy structure observed in
the debris disk, as radiation forces will tend to produce a smooth spatial
distribution of particles. The millimeter-wave emission
should then trace the location of the planetesimals in the HD\,107146 disk
\citep{Wyatt06}. The radius of the observed debris emission implies the
presence of planetesimals at an orbital radius of $\sim$97\,AU. Moreover, the
dual emission peaks in the CARMA image presents intriguing possibilities for
the presumed planetary system around HD\,107146. The dust emission peaks are
distributed nearly (but not precisely) symmetrically around the star, and may
be caused by dust trapped in mean-motion resonance with an orbiting star or 
planet. In fact, the image of HD\,107146 looks quite similar to the 
$\beta$$\sim$0.002, 3:2 resonance image of \citet{Wyatt06}. In detail, the two 
main dust peaks and the star are not co-linear, which may imply a planet of 
moderate eccentricity ($e\sim$0.5) has trapped the dust in a 3:1 resonance
\citep{Kuchnar03}. Given that the clumps are located at 97\,AU, such 
resonances would imply a planet between $\sim$45 and 75~AU.

\citet{Metchev03} obtained adaptive optics images at $\lambda$2.2\micron\
to search for companions to HD\,107146. They place an upper mass limit of 
10\,M$_J$ to any companions at 74\,AU. \citet{Apai08} report that a similar
limit of 10\,M$_J$ persists inward to $\sim$25\,AU.  Thus these observations rule out the
presence of a stellar companion at such a large radius, and suggest the 
structure in the debris disk is produced from the gravitational affects of a
planetary system.

Could a planet have formed at $\sim$45-75\,AU that produced resonant structure 
in the debris disk? \citet{Kenyon04a} have simulated the
collision growth of planets and subsequent production of debris and find that
planets on the size of $\sim$2000\,km ($\sim$0.01\,M$_\earth$) are needed to
excite the collisional cascade. The timescale to produce bodies of this size at
45-75\,AU from collisional growth of planetesimals is $\sim$50-225\,Myr
\citep{Kenyon04a} for planetesimal with low eccentricity ($e<0.02$) and
a surface density profile appropriate for the minimum mass solar nebula
\citep{Weidenschilling77}. For moderate eccentricities ($e\sim$0.02-0.04), 
the timescales are 2 to 4 times longer. Similarly, \citet{Kenyon04b} simulated
the collisional growth of objects at 40-47\,AU and found that 1000\,km sized
objects can form in 10-50\,Myr, and double in mass every 100\,Myr to 1\,Gyr.
Thus the timescale to form $\sim$2000\,km sized bodies, which are needed to
initiate the collisional cascade, is roughly consistent with the age 
estimates of HD\,107146.

However, the planet is likely much larger than 2000\,km in radius if the dust
is trapped in a resonance. \citet{Roques94} conducted numerical simulations to
show that planets more massive than $\sim$5\,M$_\earth$ ($\sim$11,000\,km in
radius assuming a volume density comparable to the Earth) are needed to
trap particles in outer mean motion resonances. The timescale to grow such
objects by collisional growth is more than 1\,Gyr at radii $>$45\,AU 
\citep{Kenyon04b}, which is substantially older than HD\,107146.

The presence of a massive planet at large radii then requires formation
mechanisms beyond collisional aggregation of planetesimals. For disk-to-star
mass ratios of $\ge$0.1 in the T~Tauri phase, marginally consistent with
observations \citep{Andrews07}, planets may form by gravitational instability
\citep[e.g.][and references therein]{Boss08}. Planets formed by this mechanism
tend to be quite massive (1\,M$_{\rm J}$ or larger).  \citet{Rafikov05} argues
that the cooling timescales are inconsistent with formation via this method
unless the planet is both massive, $\sim$5-10\,M$_J$, and distant,
$\sim$100\,AU, which is consistent with the distance and upper mass limit
described above. Another possibility is that the planet formed in the inner
regions and migrated outwards, which may trap planetesimals and dust in 
resonant orbits \citep{Wyatt03}. Such a mechanism has been used to explain the
structure and dynamics of the Kuiper Belt \citep{Malhotra93,Malhotra95}. 
Detailed modeling is needed to gain more insights on the planetary system
around HD\,107146 and if migration can explain the clumpy structure observed
in the surrounding debris disk.

In summary, the young solar analog HD\,107146 is surrounded by a clumpy debris disk
similar to those seen around Vega and $\epsilon$Eridani.  The
structure is suggestive of dynamical influence from a planet.  The
likely size and location of the presumed planet are inconsistent with
formation by purely collisional aggregation of planetesimals and
appeals to migration or formation via gravitational instability must
be made.

\acknowledgments 

We would like to thank the referee for comments which improved this work.
Support for CARMA construction was derived from the Gordon and Betty Moore
Foundation, the Kenneth T. and Eileen L. Norris Foundation, the Associates of
the California Institute of Technology, the states of California, Illinois, and
Maryland, and the National Science Foundation. Ongoing CARMA development and
operations are supported by the National Science Foundation under a cooperative
agreement, and by the CARMA partner universities.

\end{document}